\documentclass[12pt]{article} 
\usepackage{epsfig}
\def\munu{\mu_{\nu}}

\begin{document}

\hfill AS-TEXONO/10-02 \\
\hspace*{1cm} \hfill \today

\begin{center}
\large  
{\bf
Derivations of 
Atomic Ionization Effects Induced by
Neutrino Magnetic Moments\\[2ex]
}
\normalsize
H.T.~Wong\footnote{Corresponding~author $-$\\
\hspace*{1cm} Email:~htwong@phys.sinica.edu.tw;
Tel:+886-2-2789-6789;
FAX:+886-2-2788-9828.},
H.B.~Li, and
S.T.~Lin \\[2ex]
Institute of Physics, Academia Sinica, Taipei 11529, Taiwan
\end{center}
\normalsize

\begin{abstract}

A recent paper [M.B. Voloshin, Phys. Rev. Lett.
{\bf 105}, 201801 (2010)] pointed out 
that our earlier derivations
of atomic ionization cross-section due to
neutrino magnetic moments (arXiv:1001.2074v2)
involved unjustified assumptions.
We confirm and elaborate on this comment with these notes.
We caution that the results of the sum-rule approach in
this paper contradict the expected behaviour 
in atomic transitions.

\end{abstract}

Advances in low energy detectors make 
it relevant to evaluate 
atomic effects induced by
possible neutrino electromagnetic interactions
A recent paper (Ref.~\cite{voloshin}) 
observed that there are unjustified
assumptions implicit in our previous
derivations of the atomic ionization (AI) cross-section
induced by neutrino magnetic moment ($\munu$)~(Ref.~\cite{munuaiold}).
This comment is correct. 

We use the pre-defined notations in Ref.~\cite{munuaiold}
and work with positive $q^2$ for clarity.
The d$\sigma$/dT formula in Eq.10 
is due to integration of Eq.8 over d$\Omega$ which,
implicitly, is integration over $q^2$. 
However, the $q^2 \rightarrow 0$ limits
have been taken in the assignments of 
the form factors $F_a ( q^2 , T )$
and $q^2 F_b ( q^2 , T )$. 
There is an unjustified
assumption in Eq.10
that the form factors are constant 
within the integration range of
$q^2$ from 0 to $\sim 4 E_{\nu}^2$ 
(where $E_{\nu} \sim$few MeV for reactor neutrinos).
Consequently, Eq.10 as well as the
results that follow are invalid.

The $q^2$-dependent components 
of Eq.8 in the laboratory
frame can be written as:
\begin{displaymath}
\frac{ d ~ ^2 \sigma }{ dT d \Omega }
\propto
\left[ 
M E_{\nu} 
( \frac{E_{\nu}}{T} - 1 ) 
F_a ( q^2 , T ) 
+
\frac{1}{4} q^2 F_b ( q^2 , T ) \right]  ~~~ , ~~~ {\rm where} ~~~
F_a ( 0 , T ) \propto \sigma_{\gamma}  (T) ~~~ .
\end{displaymath}
There are no experimental constraints on
$F_a ( q^2 , T )$ and $F_b ( q^2 , T )$. 
It is natural to expect the 
electron mass scale ($m_e$)
plays an important role.
In the case where the form factors 
are exclusively defined
by $m_e$, such that they are suppressed at 
$q^2 > 2 m_e T$ ($\sim$(0.1~MeV)$^2$), the
AI effects would be small 
compared to the free-electron cross-section.
If, however, a higher mass scale like that 
of the atomic mass may have even a minor role
to play in the process, 
the form factors can be finite
up to $q^2 \sim E_{\nu}^2$. 
Large AI contributions are possible in this scenario and
the discussions of Ref.~\cite{munuaiold}
would still hold.

It is instructive to note how the equivalent photon
approximation approach of Ref.~\cite{munuaiold}
does produce valid results in two similar, but non-identical,
problems.
\begin{enumerate}
\item $\munu$-induced deuteron disintegration 
with solar neutrinos~\cite{gmm04} $-$
the form factors are defined by the nucleon mass scale ($\sim$GeV)
and so can be taken as constant within the range of integration up to
$q^2 \sim ($10~MeV)$^2$, so that Eq.10 remains valid.
\item Charge-induced AI processes with relativistic 
minimum ionizing particles~\cite{mipai} $-$
the kinematics involves 
an additional (1/$q^2$) weight factor.
The integral is dominated 
by contributions at $q^2 \rightarrow 0$
and insensitive to the behaviour
of the form factors at large $q^2$. 
It is adequate to describe them by the 
physical photoelectric cross-section at $q^2 = 0$.
\end{enumerate}

Ref.~\cite{voloshin} adopted a sum-rule approach to
arrive at an inclusive cross-section of $\munu$-induced
scattering with atomic electrons. This is 
given in Eq.13 as:
\begin{displaymath}
\frac{ d \sigma }{ dT } \propto
( 1 - \delta ) ~ \frac{Z}{T} ~~ ,  ~~~~~ {\rm where }  ~~~~~
\delta = \frac{ T^2 \sigma_{\gamma} (T)  }{ 4 \pi ^2 \alpha Z} 
~~~~~ {\rm and} ~~~~~
 0 < \delta <  10^{-3} ~~~.
\end{displaymath}
In atomic transitions where binding energies ($\Delta_b$)
are involved, the cross-sections are expected to have 
a sharp increase across the transition edge
from $T < \Delta_b$ to $T > \Delta_b$.
The sum-rule results, however, represent a 
continuous cross-section,
smoothed to  $< 10^{-3}$. 
In addition, the contribution 
of photoelectric cross-section $\sigma_{\gamma}$ to
the inclusive process is negative, which implies the
total cross-section actually {\it decreases} across
the transition edge. Both features contradict the
expected behaviour. The results should therefore be
taken with caution. 
We note that an alternative derivation
using Hartree-Fock techniques results in 
a cross-section resembling that for 
free electrons scattering modified by
step-functions~\cite{aihf}. 

We are grateful to Prof. Voloshin for pointing out our
error in Ref.~\cite{munuaiold}, and for the 
subsequent stimulating and in-depth discussions.

\end{document}